\documentclass[pre,twocolumn,superscriptaddress]{revtex4}
\usepackage{amsmath,amssymb,dsfont,graphicx,amssymb,xcolor,mathrsfs,cjhebrew}
\usepackage[caption=false]{subfig}
\usepackage{grffile}

\newcommand{\kk}{\bar{k}}
\DeclareMathOperator{\dd}{\mathrm{d}}

\DeclareMathAlphabet{\mathpzc}{OT1}{pzc}{m}{it}
\newcommand{\numberset}{\mathbb}

\newcommand{\R}{\numberset{R}}

\begin{document}

\title{Dynamic interdependence and competition in multilayer networks}

\author{Michael M. Danziger}
\email{michael.danziger@biu.ac.il}
\affiliation{Department of Physics, Bar-Ilan University, Ramat Gan, Israel}

\author{Ivan Bonamassa}
\email{ivan.bms.2011@gmail.com}
\affiliation{Department of Physics, Bar-Ilan University, Ramat Gan, Israel}

\author{Stefano Boccaletti}
\affiliation{CNR- Institute of Complex Systems, Via Madonna del Piano 10, 50019 Sesto Fiorentino, Florence, Italy}
\affiliation{The Italian Embassy in Israel, 25 Hamered Street, 68125 Tel Aviv, Israel}

\author{Shlomo Havlin}
\affiliation{Department of Physics, Bar-Ilan University, Ramat Gan, Israel}

\date{\today}
\begin{abstract}
From critical infrastructure, to physiology and the human brain, complex systems rarely occur in isolation.  
Instead, the functioning of nodes in one system often promotes or suppresses the functioning of nodes in another.
Despite advances in structural interdependence, modeling interdependence and other interactions between dynamic systems has proven elusive. 
Here we define a broadly applicable dynamic dependency link and develop a general framework for interdependent and competitive interactions between general dynamic systems.
We apply our framework to studying interdependent and competitive synchronization in multi-layer oscillator networks and cooperative/competitive contagions in an epidemic model.
Using a mean-field theory which we verify numerically, we find explosive transitions and rich behavior which is absent in percolation models including hysteresis, multi-stability and chaos. 
The framework presented here provides a powerful new way to model and understand many of the interacting complex systems which surround us.
\end{abstract}

\maketitle

Many real-world complex systems include macroscopic subsystems which influence one another.
This feature arises, for example, in competing or mutually reinforcing neural populations in the brain \cite{fox-pnas2005,stumpf-epl2010,misic-neuron2015}, opinion dynamics among social groups \cite{halu-epl2013}, and elsewhere \cite{angeli-pnas2004,helbing-nature2013}.
It is therefore important to understand the possible consequences that different types of inter-system interactions might have. 
In 2010, substantial progress was made when the theory of percolation on interdependent networks was introduced \cite{buldyrev-nature2010}.
This model showed that when nodes in one network depend on nodes in another to function, catastrophic cascades of failures and abrupt structural transitions arise, as observed in real-world systems \cite{buldyrev-nature2010,leichtdsouza2009,kivela-jcomnets2014,boccaletti-physicsreports2014,danziger-collection2016}.\\
\indent
However interdependent percolation is limited to systems where functionality is determined exclusively by connectivity: either to the largest connected component \cite{buldyrev-nature2010}, backbone \cite{danziger-newjphysics2015} or a set of source nodes \cite{min-pre2014}. It thus provides only a partial understanding of real-world systems, where the network serves as the base upon which a dynamic process occurs~\cite{boccaletti-physicsreports2002,barzel-naturephysics2013}.

Here, we propose a general framework for modelling interactions between dynamical systems.
Two fundamental and ubiquitous ways in which nodes in one system can influence nodes in another one are \textit{interdependency} or cooperation, as in critical infrastructures \cite{rosato-criticalinf2008,shekhtman-chaos2016,danziger-collection2016} or among financial networks \cite{kenett-mindsociety2015,brummitt-pre2015}, and \textit{competition} or antagonism, which is common in ecological systems \cite{hibbing-naturereviews2009,coyte-scienceadvances2015}, social networks \cite{halu-epl2013}, or in the human brain~\cite{fox-pnas2005,nicosia-prl2017}. 
It is not uncommon to find interdependent and competitive interactions simultaneously,  in predator-prey relationships in ecological systems \cite{blasius-nature1999}, in binocular rivalry in the brain \cite{fries-pnas1997}, or even in phenomena like ``frenemies'' and ``coopetition'' in social systems~\cite{lee-pubadmin2012}.
Special cases of  competitive interactions between networks have been studied, but without a general framework or ability to uncover universal patterns between systems~\cite{masuda-jtheoreticalbio2006,gomezgardenes-royalsocietya2015,valdez-jstatmech2016}.

We model the cross-system interaction by multiplying the \textit{coupling strength} of a node  to its neighbors in one network by a function of the \textit{local order} of a node in another network.
If the function is increasing, the potential for local order of the two nodes is positively correlated, reflecting an interdependent interaction; if it is decreasing, then the potential for local order is anti-correlated, reflecting a competitive coupling.
Because local order often reflects the instantaneous local functionality and can be meaningfully defined for a wide range of complex systems, this framework can capture an unprecedented variety of inter-system interactions.

After presenting the general equations for dynamical dependence, we examine in detail a system of two networks of Kuramoto oscillators and a system of two susceptible-infected-susceptible (SIS) epidemic processes, with different combinations of competitive and interdependent interactions between them.
We find that under an interdependent interaction, the system exhibits behavior familiar from interdependent networks: abrupt transitions from order to disorder and cascade plateaus at criticality.
Furthermore, because of the added richness of the dynamical models, new features such as a forward transition (and hysteresis), and a metastable region are observed.
Similarly under a competitive interaction, we find coexistence, hysteresis and multi-stability.
When the two types of interactions are asymmetrically implemented, we observe novel oscillatory states and chaotic attractors.
Because the new interactions are expressed via local order terms, we are able to perform a mean-field approximation of the exact equations, which are solved numerically and verified against extensive GPU-accelerated simulations on large synthetic networks.
\begin{figure*}
    \centering
    \includegraphics[width=0.9\textwidth]{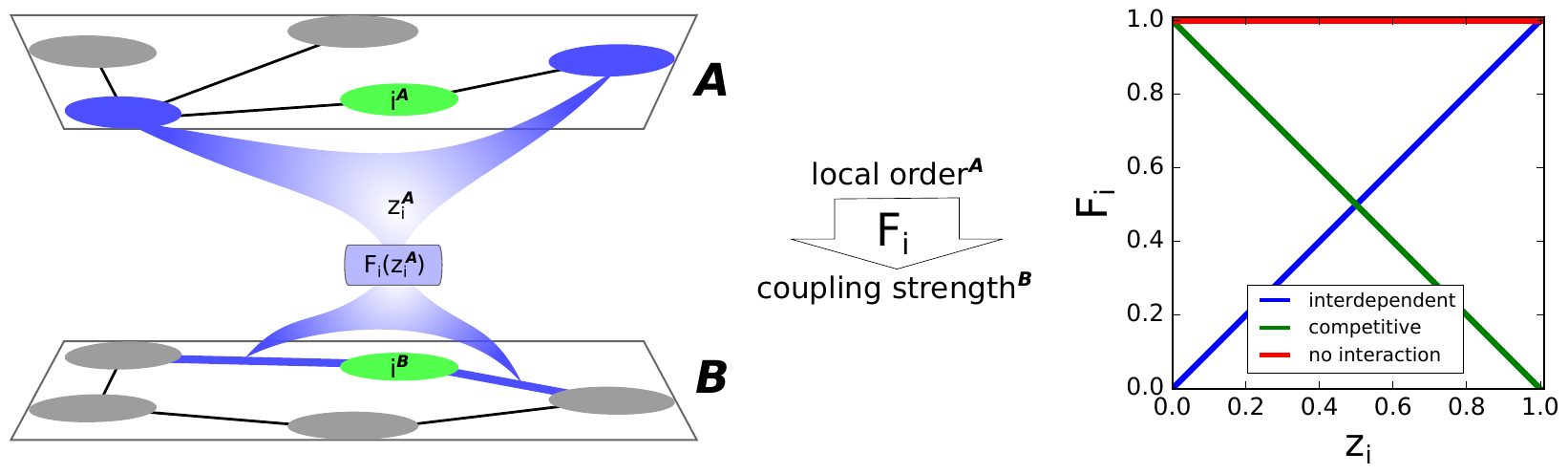}
    \caption{{\bf Dynamic interdependence and competition} (color online). Left panel: the local order $\mathpzc{z}_i^{\mathrm{A}}$ of node $i$ in network $\mathrm{A}$ is determined by the state of its neighbors. This then modifies the effective strengths of the incoming links of node $i$ in network $\mathrm{B}$, according to a function $F_i^{\mathrm{A}\to\mathrm{B}}$ of $\mathpzc{z}_i^\mathrm{A}$ which can reflect cooperative, antagonistic or other interactions. Note that there are typically interactions in the opposite direction as well, i.e. $F_i^{\mathrm{B} \to \mathrm{A}}$, which have not been drawn here for the sake of clarity. Right panel: summary of the dynamical interaction strategies considered in the main text. We will adopt linear coupling functions, randomly distributed among nodes, modeling hence interdependence (blue), competition (green), and no interactions (red).}
    \label{fig:fig1}
\end{figure*}

\begin{figure*}
\newcommand{\figonepanelh}{6cm}
    \centering
    \subfloat[]{\includegraphics[height=\figonepanelh]{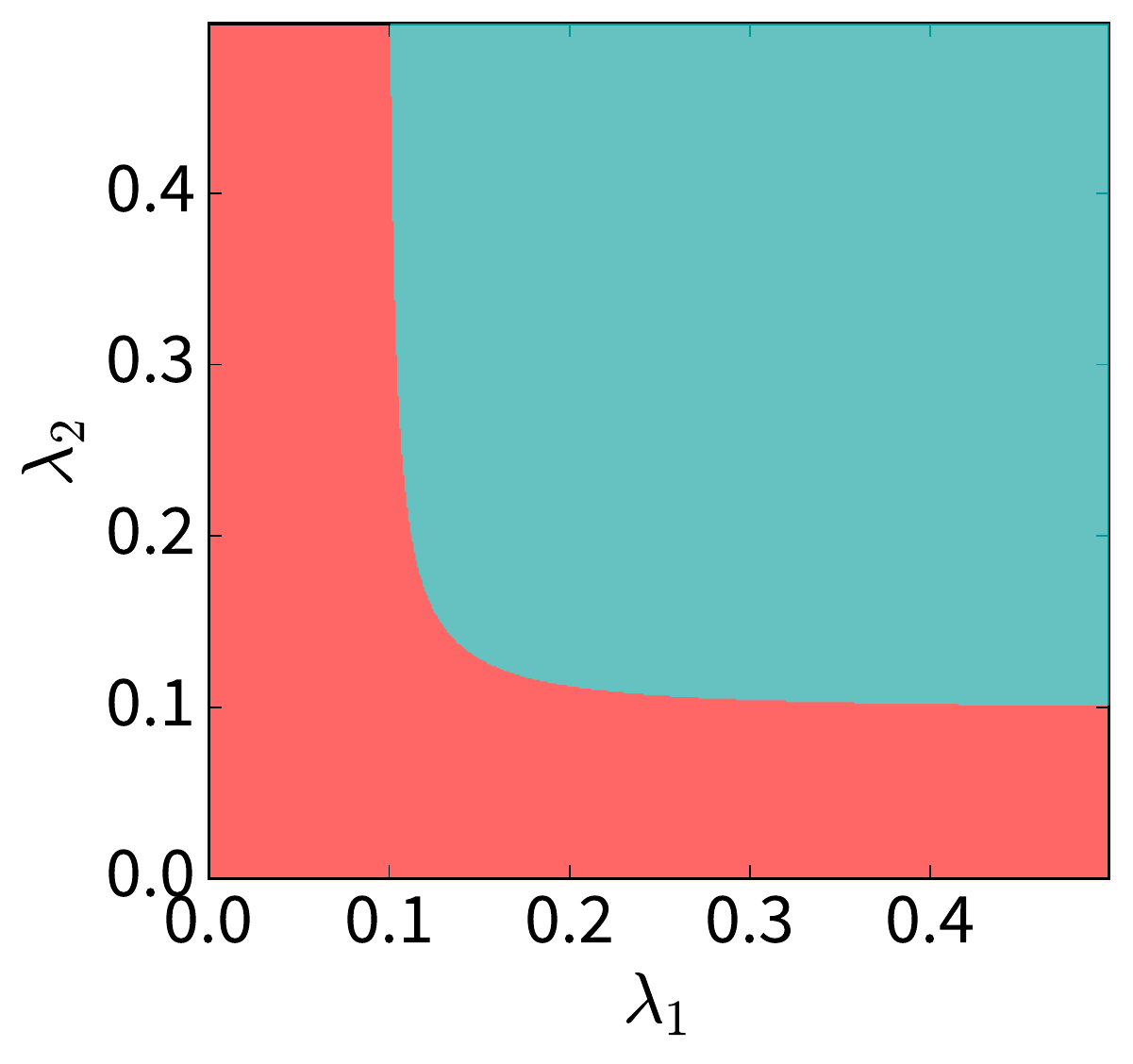}\label{subfig:id_theory_f1}}
    \subfloat[]{\includegraphics[height=\figonepanelh]{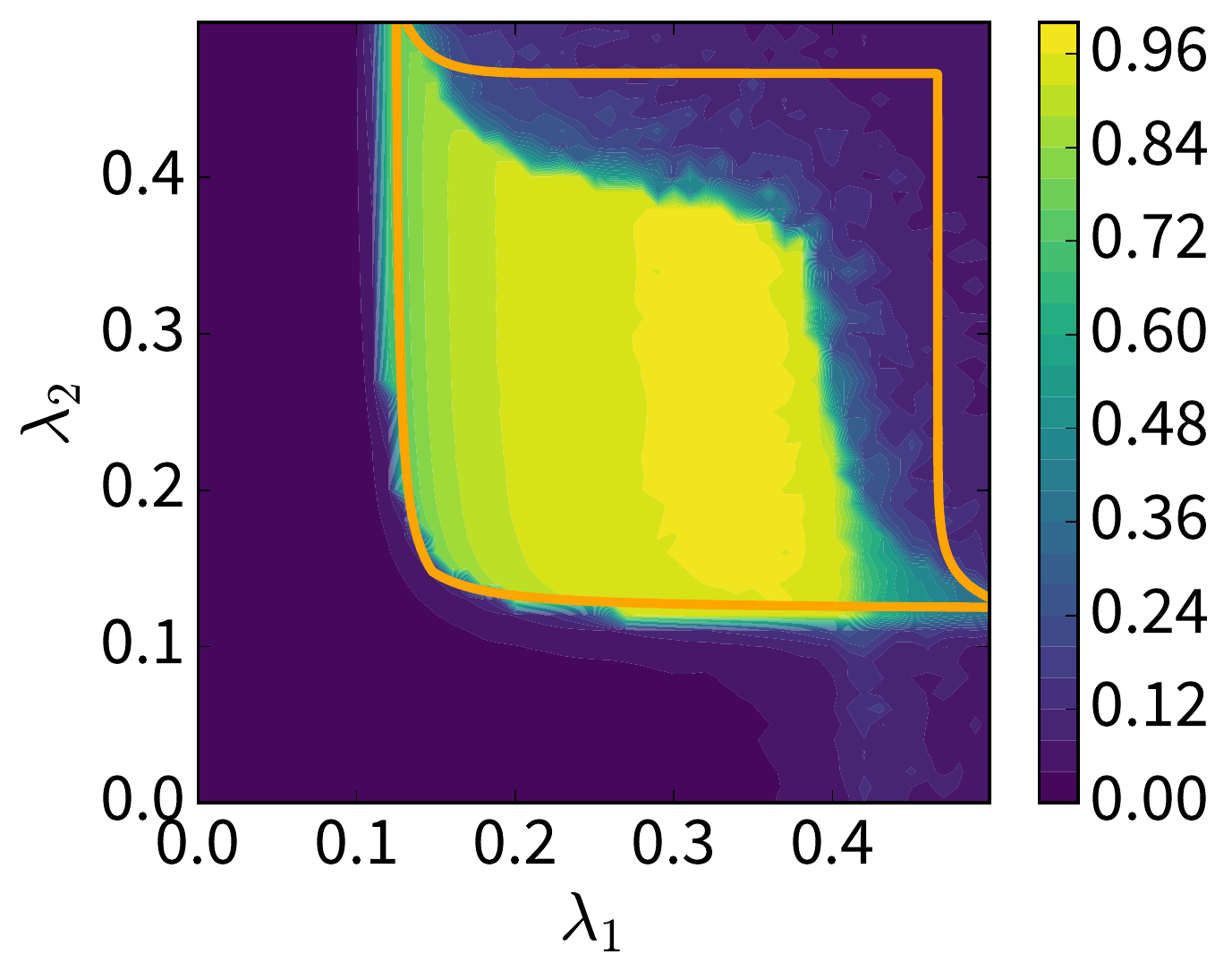}\label{subfig:id_metastable_f1}}\\
    \subfloat[]{\includegraphics[height=\figonepanelh]{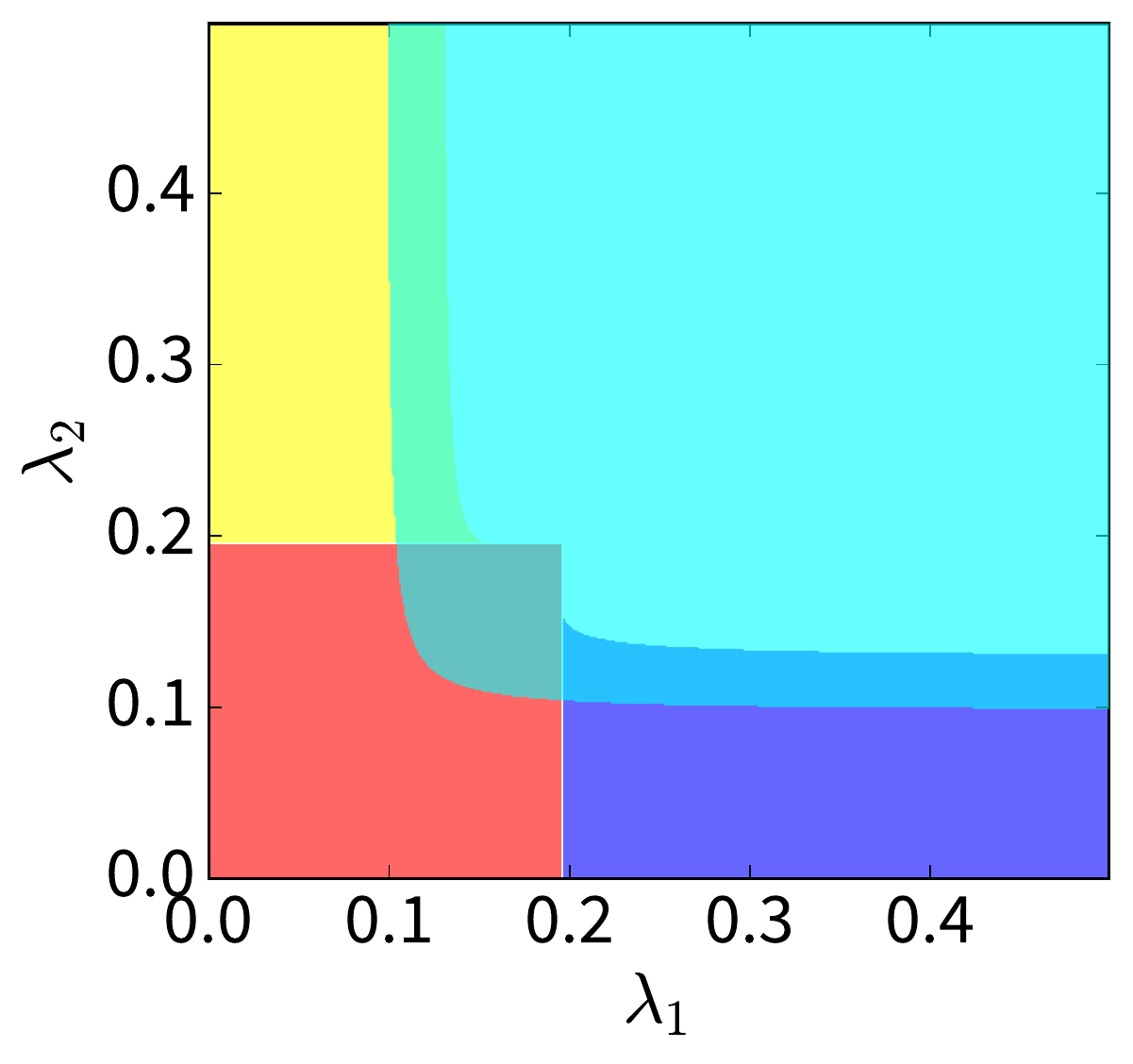}\label{subfig:id_theory_f05}}
    \subfloat[]{\includegraphics[height=\figonepanelh]{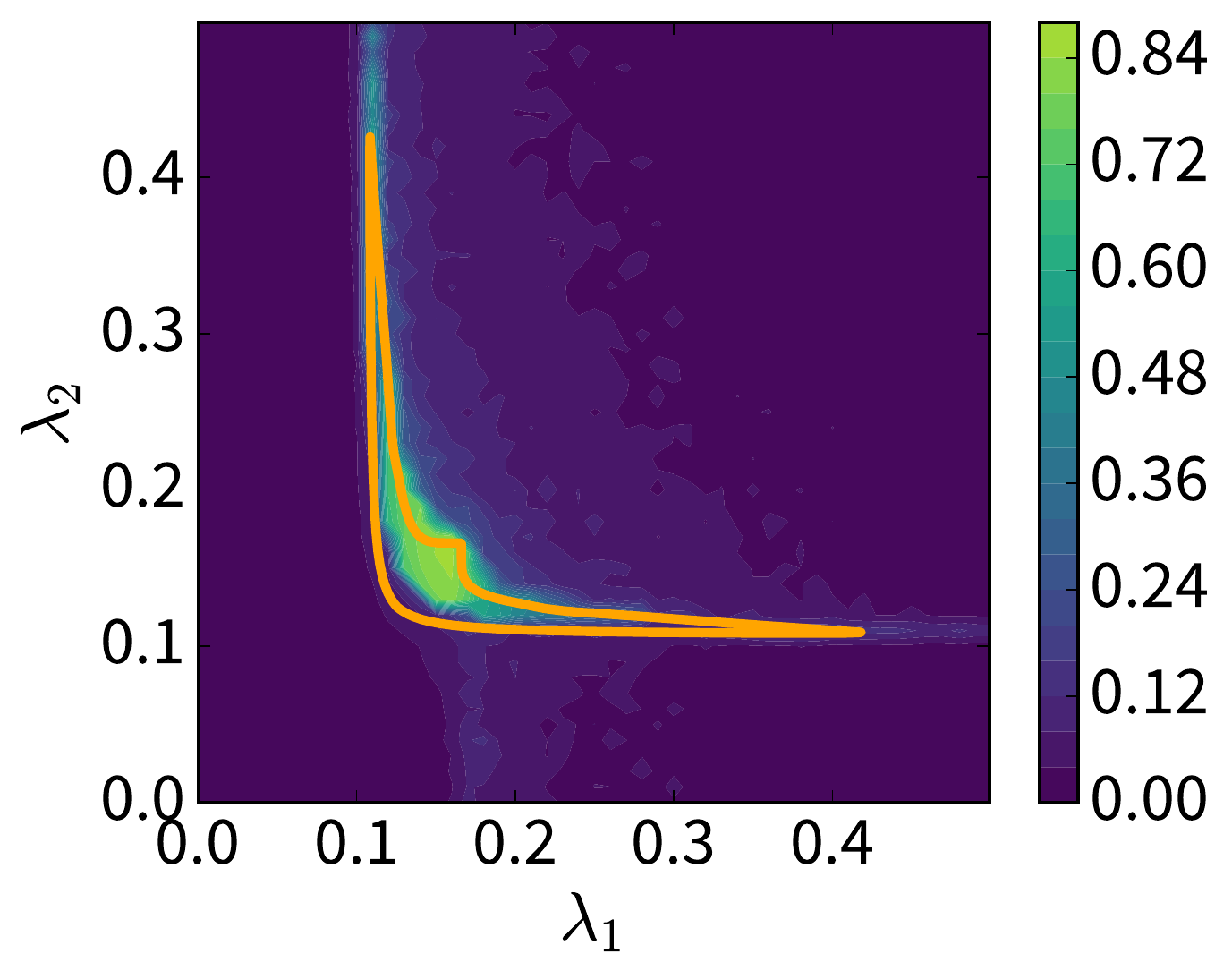}\label{subfig:id_metastable_f05}}
    \caption{\textbf{Interdependent synchronization} (color online).
    \textbf{(a)} Predicted phase space for fully ($f = 1$) interdependent networks. There are two regions: red where no network is synchronized and cyan-on-red where both synchronized and desynchronized solutions are stable. \textbf{(b)} Difference in final synchronization depending on initial condition. The yellow region represents the metastable region where the explosive transition takes place. The orange line is the predicted extent of the metastable region from the mean field theory, taking into account that there are characteristic fluctuations which can spontaneously jump solutions of size $\Delta R \approx 0.2$. \textbf{(c)} Predicted phase space for partially ($f = 0.5$) interdependent networks.  In addition to the phases in \textbf{(a)}, we have network 1 (2) only marked in blue (yellow) which coexists with the both-synchronized solution where the cyan overlaps.  Additionally, we have a cyan-only region where the zero-solution is not stable at all. \textbf{(d)} The metastable region and prediction for the partially interdependent case.}
    \label{fig:interdep}
\end{figure*}
\section{Model}

{
We begin by considering isolated dynamical systems composed of $N$ nodes which evolve according to
\begin{equation}\label{eq:0}
    \dot{x}_i = g(x_i) + \lambda \sum_{j=1}^N A_{ij} h(x_i,x_j)
\end{equation}
where $x_i$ is the dynamic state of node $i$, $g$ and $h$ are scalar-valued functions of self-dynamics and pairwise interactions, respectively, $\lambda$ is the coupling strength and the network of interactions is represented by $A_{ij}$ which is equal to one if nodes are connected and zero otherwise.
An extremely wide range of physical models can be described by equations of this form, including oscillators, epidemic processes, opinion models and many others.
The question that we raise here is, how can a dependency relationship be defined between systems such as these?
Specifically, we want to include the effects of local functionality in one system promoting or suppressing the onset of functionality in another system.
To address this lack, we propose to define a link between systems via multiplication of the coupling strength $\lambda$ of a node in one network by a function of the \textit{local order} $z_i(t)$ around a node in another network. 
Local order is defined as the weighted average of the ordered state of the neighbors of $i$, with the precise definition dependent on the model in question.  In general we replace:
\begin{equation}
    \lambda_B \rightarrow \lambda_B F_i^{A\to B}(t)
\end{equation}
where $F_i$ is a function of the local order parameter $z_i(t)$ of node $i$ in layer $A$.  Because the numbering of the nodes is arbitrary, we assume that node $i$ in layer $A$ affects node $i$ in layer $B$.  As seen in Fig. \ref{fig:fig1}, a suitable choice of $F_i$ can represent an interdependent or competitive interaction.
We thus consider
\begin{equation}\label{eq:interaction}
F_i^{A\to B}(t) = 
\begin{cases}
\left|z_i^{A}\right|(t) & \mbox{interdependent},\\
1 - \left|z_i^{A}\right|(t) & \mbox{competitive},\\
1 & \mbox{decoupled}
\end{cases}
\end{equation}
and leave more exotic interactions for future study.

In this manner, we can describe the evolution of the dynamic state in layer $\sigma$ of an $M$-layer ensemble of interdependent, competitive or mixed interacting dynamical systems with the equations:
\begin{equation}\label{eq:model}
\dot{x}_i^\sigma = g(x_i^{\sigma})+\lambda_\sigma\prod_{\mu=1}^M F_i^{\mu\to \sigma}\sum_{j=1}^{N} A_{ij}^{(\sigma)}h\big(x_i^\sigma,x_j^\sigma \big),
\end{equation}
where we assume, for simplicity, that the $g$ and $h$ functions are the same in each layer, i.e., the same process is taking place in each layer.
We further assume that the time scale of the processes is the same.
The product in Eq. \eqref{eq:model} reflects the assumption that if the interaction term $F_i^{\mu\to \sigma}$ goes to zero in any of the layers, it suppresses the coupling of node $i$ in every layer, reflecting mutual interdependence (or competition).

We can thus consider the $F_i^{\mu\to \sigma}$ terms as elements of a supra-adjacency matrix $F_i(t)$ describing the interactions between layers at node $i$, which for the case of two interacting networks $A$ and $B$ would be represented as:
\begin{displaymath}
F_i(t)=\begin{pmatrix}
1 & F_i^{B\to A}(t)\\
F_i^{A\to B}(t) & 1
\end{pmatrix}
\end{displaymath}
where we have assumed that there are no self-interactions.

Because $F_i$ is determined entirely by the local order parameter, it is straightforward to analyze the system with a mean-field approach by simply replacing $z_i$ with $\mathpzc{Z}$, the corresponding global order parameter.  
For many systems described by Eq. \eqref{eq:0}, the sum on neighbors can be rewritten as a function of the local order parameter, and this is used as the basis for a mean-field theory.  
Because we also define the interaction in terms of local order, the mean-field approach simultaneously solves the intra-layer and inter-layer dynamics.
For two networks, we can then determine a general self-consistent equation for the global order parameter:
\begin{equation}\label{eq:genselfconst}
    \mathpzc{Z}_\sigma = \int \frac{k P_\sigma(k)}{\langle k\rangle_\sigma} G_\sigma(\lambda_\sigma \mathpzc{Z}_\sigma F^{\mu\to\sigma}, k)\mathrm{d}k,
\end{equation}
where $F^{\mu\to\sigma}$ is equal to the mean-field interaction term: $\mathpzc{Z}_\mu$ for interdependent and $1 - \mathpzc{Z}_\mu$ for competitive and the $G$ function is a dynamics-dependent function based on the mean-field solution of the single layer case. One feature of this function is that it always has a multiplicative factor of $\mathpzc{Z}_\sigma$ and therefore a zero solution, which may or may not be stable.
This equation holds when all of the nodes are coupled.
In the methods section, we provide the generalized equation for the case where only a fraction $f<1$ are coupled and the remaining nodes are decoupled.
Thus for any mean-field solvable dynamical system, our framework enables study of the evolution of the coherent state under any combination of competitive and interdependent links.
\begin{figure*}
    \centering
    \begin{minipage}{0.35\textwidth}
    \subfloat[]{\includegraphics[width=\textwidth]{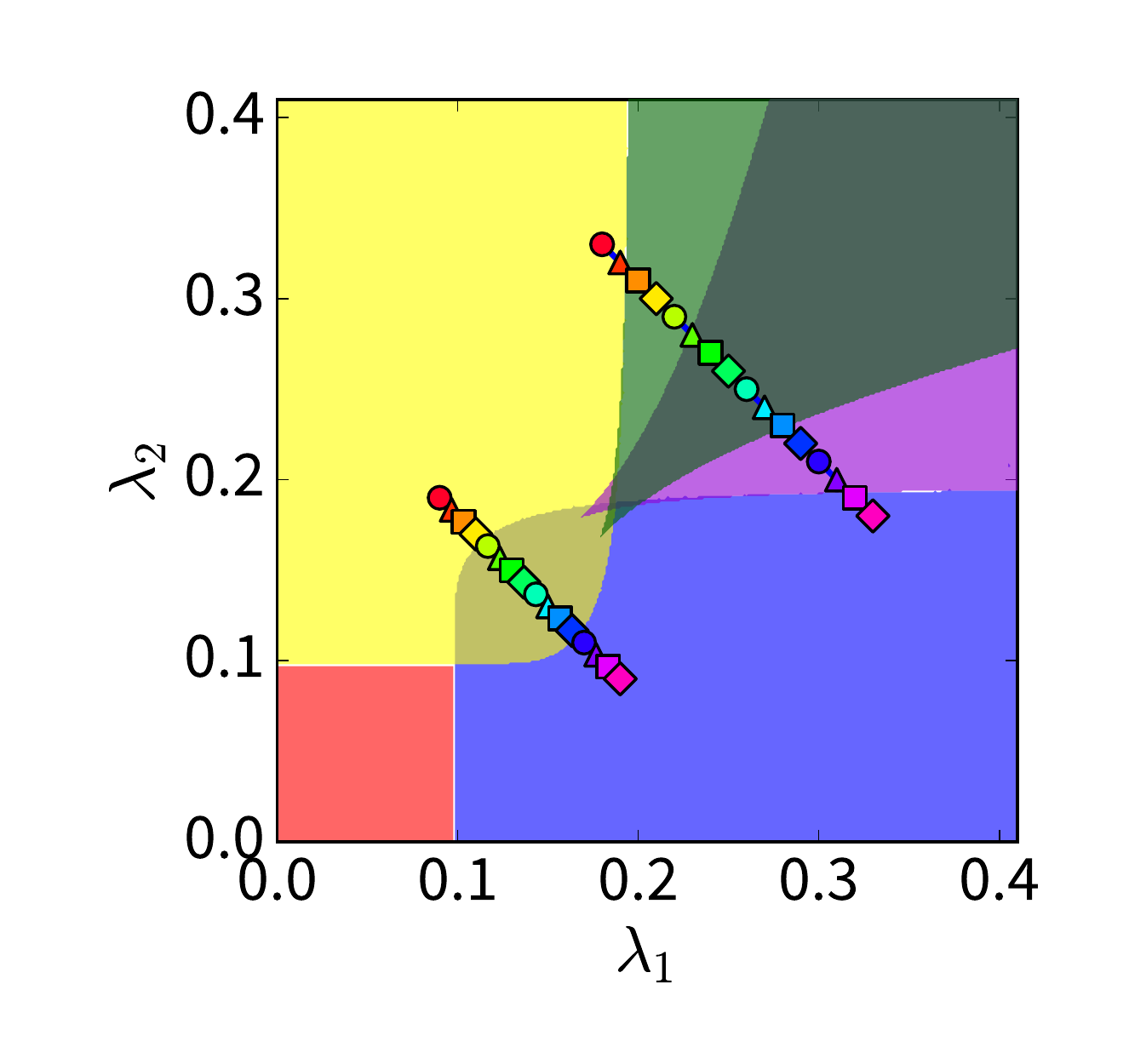}}
    \end{minipage}
    \begin{minipage}{0.62\textwidth}
    \subfloat[]{\includegraphics[width=\textwidth]{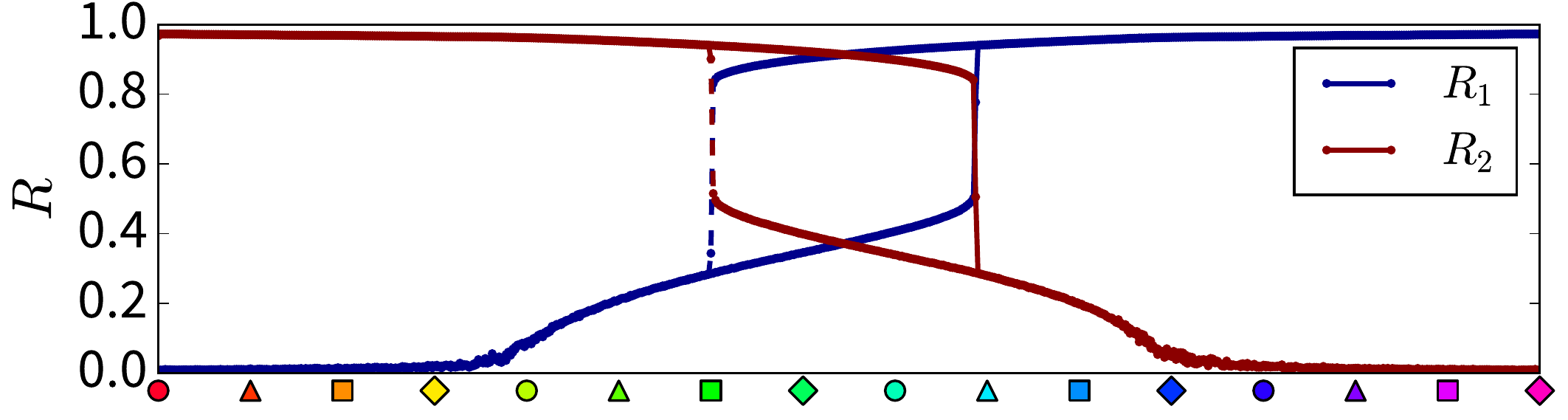}}\\
    \subfloat[]{\includegraphics[width=\textwidth]{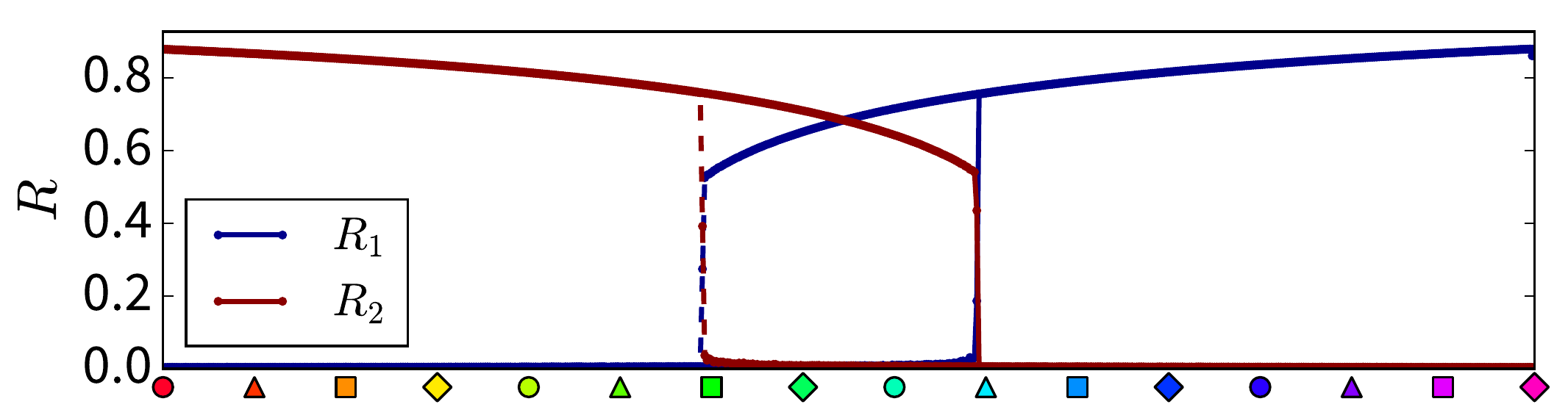}}
    \end{minipage}
    \caption{{\bf Competitive synchronization} (color online). Complex hysteresis regions in partially competitive networks.
    (a) The phase space for a system of two partially ($f=0.5$) competitive networks.
    The upper and lower lines correspond to panels (b) and (c), respectively.
    We find that, in fact, there are two transitions, a continuous transition and a discontinuous transition as the non-competing nodes begin to synchronize, even as the competing nodes are suppressed from synchronizing due to the interaction. 
    Simulation system size is $N=2^{17}$.}
    \label{fig:hysteresis}
\end{figure*}
In the following, we will demonstrate how this new type of cross-layer dynamic link impacts the onset of macroscopic order in two archetypal systems: Kuramoto oscillators and reversible (SIS) epidemics.  We examine these systems in three configurations of cross-layer links: (1) interdependent both ways, (2) competitive both ways and (3) one way interdependent and one way competitive (henceforth referred to as ``hybrid''). All three cases have real-world motivations and are studied here with full coupling ($f=1$) and partial coupling ($f<1$).

\begin{figure*}
\subfloat[]{\includegraphics[width=0.32\textwidth]{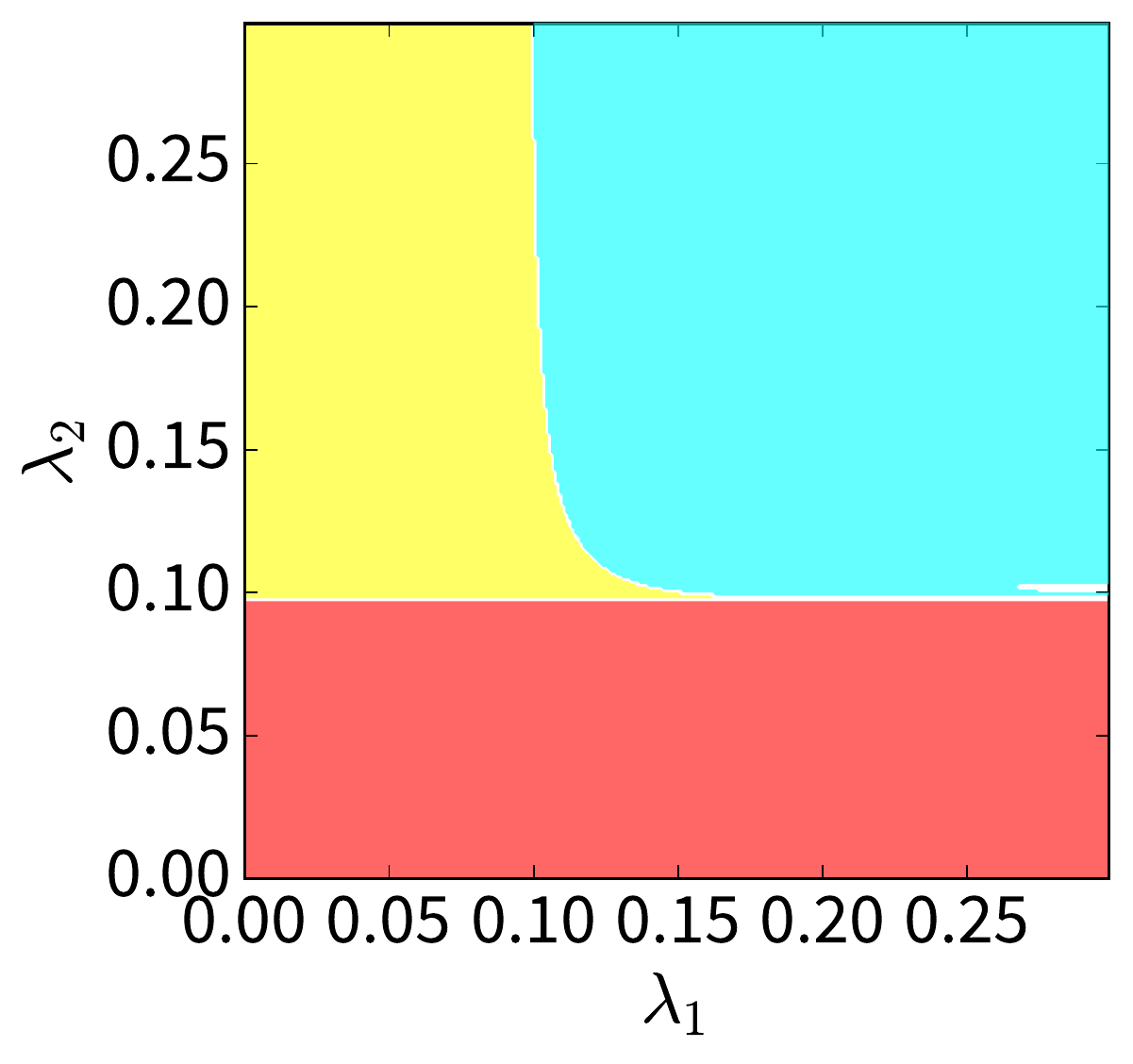}}
\subfloat[]{\includegraphics[width=0.32\textwidth]{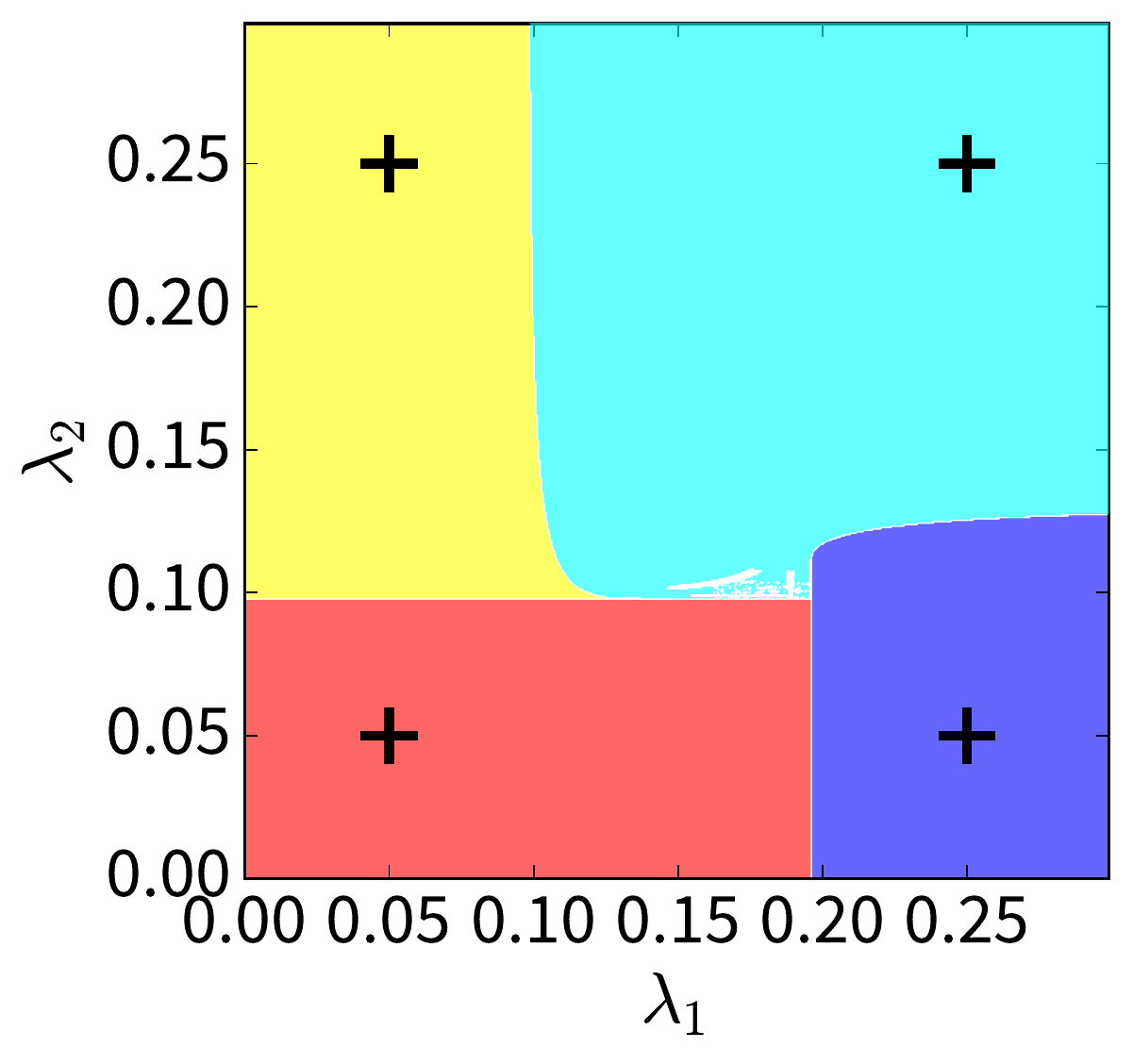}}
\subfloat[]{\includegraphics[width=0.32\textwidth]{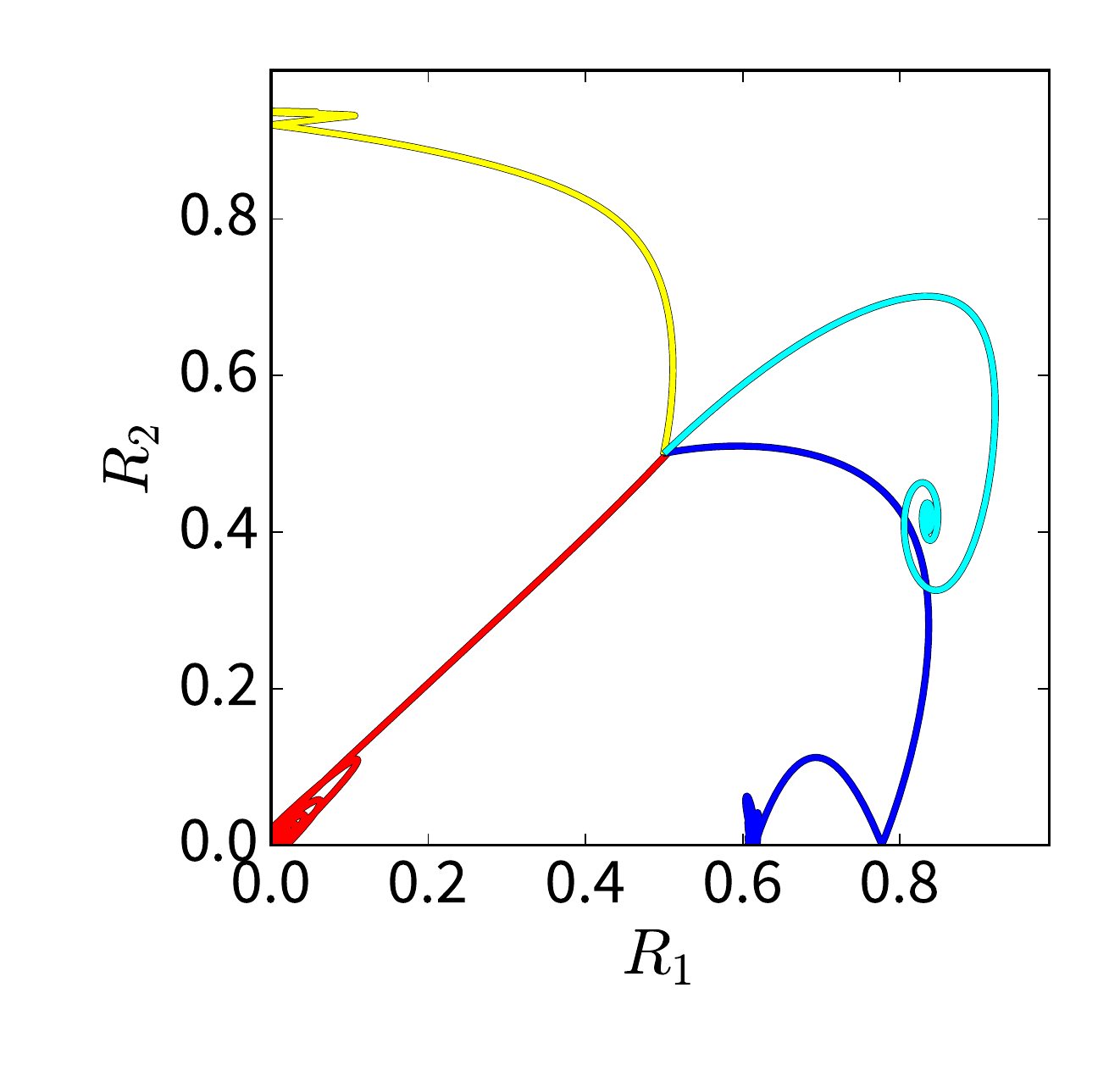}}\\
\subfloat[]{\includegraphics[height=5.5cm]{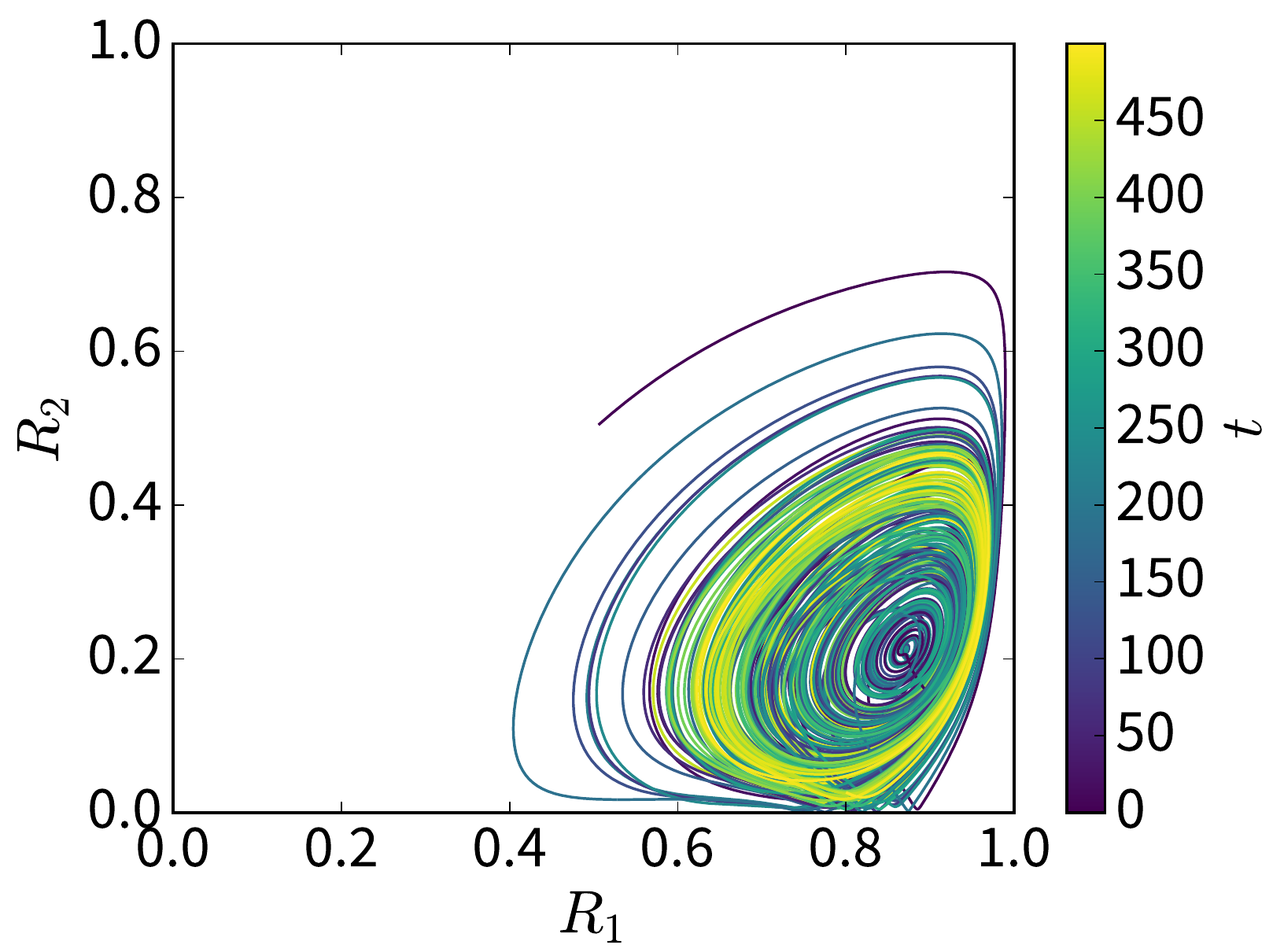}}
\subfloat[]{\includegraphics[height=5.5cm]{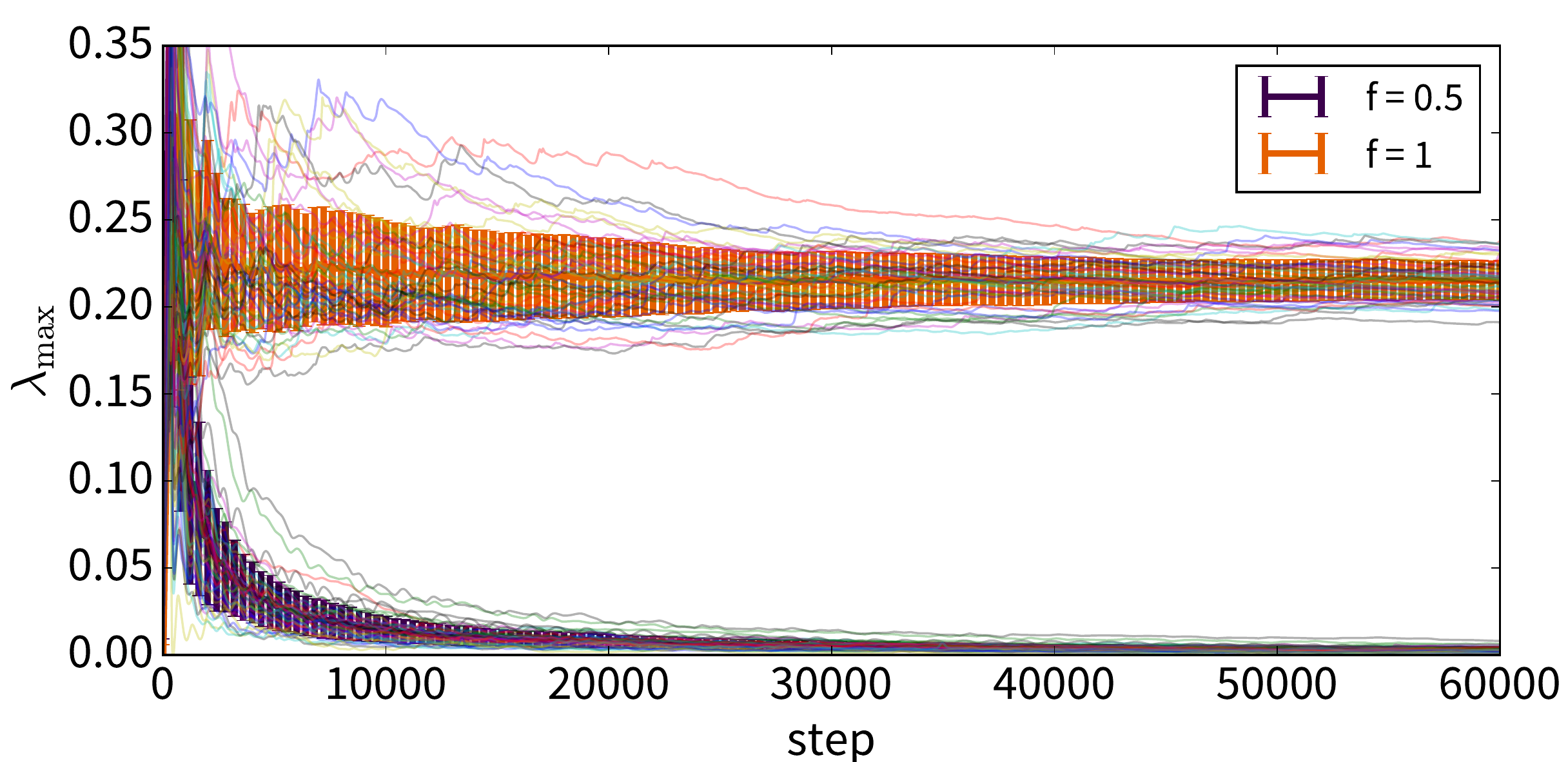}}
\caption{{\bf Hybrid synchronization} (color online). \textbf{(a)} Phase space for fully interacting case ($f=1$).
Because network 1 depends on network 2 while suppressing network 2, there is no state that is stable for it alone.
\textbf{(b)} Partially interacting case ($f=0.5$).
\textbf{(c)} Trajectories through state-space for systems corresponding to the three locations marked in panel \textbf{(b)}.  Note the collapsing limit-cycle for the blue curve (corresponding to the cyan region), where both networks are synchronized.  For each point, the system is initialized at the same synchronization level and integrated for $t=500$ for ER networks with $\kk =12$ and $N = 2^{21} \approx 2M$ nodes per layer.
Macroscopic chaos for fully hybrid synchronization (color online). 
\textbf{(d)} Sample of a typical chaotic trajectory. In contrast to the partial ($f=0.5$) hybrid case described in Fig.~\ref{fig:hybrid}, i.e. where there are sufficient non-interacting nodes to maintain a non-zero global synchronization state, in fully hybridly coupled systems the macroscopic state does not converge to its mean-field fixed point, but exhibits instead a chaotic behavior.
\textbf{(e)} To verify the chaotic behavior, we measure the largest Lyapunov exponent $\lambda_{\rm max}$ for a system with $N=2^{15}\approx 10^5$ nodes in each layer, using the standard method as explained in e.g. \cite{sadilek-scireps2015}. Averaging over 50 runs, we find $\lambda_{\rm max} = 0.21 \pm 0.01$ for $f=1$ while $f=0$ is decisively non-chaotic with $\lambda_{\rm max} < 0.004$.
}\label{fig:hybrid}
\end{figure*}

\section{Interdependent and competitive synchronization}
Synchronization is a common feature of diverse physical systems.  
It has been observed that competitive and cooperative interactions between neural populations play key roles in vision \cite{fries-pnas1997} and elsewhere in the brain \cite{fox-pnas2005}.

By convention, we refer to the dynamic phase of oscillator $i$ as $\theta_i$, take $g(\theta_i) = \omega_i$ as the constant function mapping each node to its natural frequency $\omega_i$, and $h(\theta_i,\theta_j) = \sin(\theta_j - \theta_i)$, we obtain from Eq. \eqref{eq:model}, for the case of two networks:
\begin{equation}
\dot{\theta}_i^\sigma=\omega_i^\sigma+\lambda_\sigma F_i^{\mu\to\sigma}\sum_{j=1}^N A_{ij}^{(\sigma)}\sin(\theta_j^\sigma-\theta_i^\sigma),\label{eq:multiKura}
\end{equation}
where $F_i$ is defined as in Eq. \eqref{eq:interaction} and the local order parameter $z_i$ is defined as:
\begin{equation}\label{eq:localOP}
z_i(t) = r_i(t) e^{i\psi_i} =  \frac{1}{k_i} \sum_j A_{ij} e^{i(\theta_j)}
\end{equation}
where $\psi_i$ is the instantaneous average phase of the neighbors of $i$.  

For the fully interdependent case, we have: 
\begin{equation}
\dot{\theta}_i^\sigma=\omega_i^\sigma-\lambda_\sigma r_i^\mu r_i^\sigma k_i^\sigma \sin(\theta_i^\sigma-\psi_i^\sigma),\label{eq:interKura}
\end{equation}
where we have rewritten the interaction term using the local order (see Methods for details). This can be reduced to $N$ decoupled equations using a mean-field approximation $r_i^\mu=R^\mu = \lvert \sum\exp(i [\theta_i - \Psi]) \rvert$ for global average phase $\Psi$ and moving to the rotating frame by letting $\Delta\theta = \theta_i - \Psi$:
\begin{equation}\label{eq:interKura2}
\dot{\Delta\theta}_i^\sigma=\omega_i-\lambda_\sigma k_i^{\sigma}R^{\mu}R^{\sigma}\sin\Delta\theta_i^\sigma.
\end{equation}
Using a continuum approximation we then obtain a self-consistent integral equation for the global order in both networks which defines the fixed points, stability and phase flow for this system, as described in the methods section.
This leads to a self-consistent equation of the form of Eq. \eqref{eq:genselfconst} with $G(x,k) = x \int_{|\omega|<x} d\omega g_{\rm freq}(\omega) \sqrt{1 - \frac{\omega}{xk}} $, the familiar equation from the self-consistent solution following Kuramoto~\cite{kuramoto-proceedings1975}.
We note that the special case of two adaptively coupled Kuramoto networks with $\lambda^\sigma = \lambda^\mu$ has been considered previously in the context of explosive synchronization \cite{filatrella-pre2007,zhang-prl2015,danziger-chaos2016}.

We find that, similar to the mutual giant component of interdependent percolation, the global synchronization level undergoes discontinuous  backward transition as the coupling strength is decreased.
An interesting and significant departure from the percolation models is the existence of a novel forward transition from desynchronized to synchronized state.
When $f=1$ the zero solution is always nominally stable but becomes unstable only due to the existence of a nearby saddle-point (Fig. \ref{fig:interdep}a and Supp. Fig. 1-2) and fluctuations of the size of the basin of stability~\cite{zou-prl2014}.
However, for $f<1$, as the coupling strength is increased, there is a point at which the desynchronized solution ceases to be stable and the system spontaneously jumps to the synchronized branch (Fig. \ref{fig:interdep}c and Supp. Fig. 3-4).  
This forward transition does not exist in percolation models and its existence allows us to delineate a clear metastable region which we can roughly predict by assuming a characteristic fluctuation size (Fig. \ref{fig:interdep}b,d).
Metastable synchronization is absent in the standard Kuramoto model but has been observed, for example, in Josephson junctions~\cite{barbara-prl1999,filatrella-pre2007}.

Turning to the case of competitive synchronization, which is relevant for competing neural populations~\cite{fox-pnas2005}, in particular the visual processing of optical illusions \cite{eagleman-naturereviews2001},
we can follow the same steps as above but with $F_i = 1 - z_i$ to obtain:
\begin{equation}\label{eq:compKura2}
\dot{\Delta\theta}_i^\sigma=\omega_i-\lambda_\sigma k_i^{\sigma}(1 - R_{\mu})R_{\sigma}\sin\Delta\theta_i^\sigma.
\end{equation}
This system shows pronounced metastability without coexistence for $f=1$ (Supp. Fig. 5-7) and particularly interesting phenomena when $f<1$.
In addition to the simple states of desynchronization or total domination by one network or the other, we find chimera-like states in which the decoupled nodes synchronize while the competitively coupled nodes remain desynchronized leading to a global synchronization level bounded by $f$, the fraction of nodes with the competitive interaction.  
We find that the system can transition continuously between the state where only one network synchronizes to the state where one dominates but the other has chimera-like partial synchronization and then discontinuously as the fully synchronized network drops to partial synchronization and the partially synchronized network becomes fully synchronized (Fig. \ref{fig:hysteresis}).  We further find multistability of up to four solutions for a small region of the phase space (see Supp. Fig. 8-9).  These features demonstrate the rich synchronization patterns that can be caused by competing networks.

Finally, we consider the hybrid case, in which the synchronization of nodes in network $A$ promote the onset of synchronization of nodes in network $B$, even as that onset of synchronization in network $B$ suppresses the ability of nodes in $A$ to  synchronize. This type of behavior is observed in binocular rivalry, in which neurons associated with the dominant eye synchronize more strongly when the weak eye is stimulated, but the weak eye synchronizes less strongly when the dominant eye is stimulated~\cite{fries-pnas1997}.
In this case network $A$ will be described by Eq. \eqref{eq:compKura2} while network $B$ will be described by Eq.  \eqref{eq:interKura2}. In such a case, we find convergent oscillatory solutions when $f<1$ (Fig. \ref{fig:hybrid}a-c) and chaotic attractors when $f=1$ (Fig. \ref{fig:hybrid}d,e). The source of the chaotic behavior is related to the fact that when the order parameter oscillates, it causes the effective coupling strength to also oscillate and oscillatory coupling strengths have been shown to lead to the onset of chaos~\cite{so-chaos2011}.
The figures here have been derived assuming uniformly distributed natural frequencies.  Results for Lorentzian frequency distribution are shown in Supp. Fig. 10.

A unique advantage of our model is that, because the coupling between the networks is on the level of \textit{order} and not of \textit{phase}, we are now able to model the cooperative onset of synchronization at different frequency bands.
This is significant in light of the complex interactions between neural populations of different frequencies~\cite{gans-prl2009,lowet-ploscompbio2015} which cannot be captured by existing multilayer models 
and the potential for networks of synchronizing oscillators to represent learning tasks \cite{reichert-preprint2013}, modeling fundamentally multi-frequency phenomena such as hearing~\cite{wang-frontiersofphysics2016} and physiological synchronization between organ systems in the body which may be very different from one another \cite{amir-naturecomm2012,ivanov-newjphysics2016}.
These phenomena may be more naturally modelled using the idea of order affecting order rather than as oscillators representing different entities summing their phases directly.
Mathematical treatment of multimodal interdependent synchronization is presented in Supp. Sec. S3.

\begin{figure*}
    \centering
    \subfloat[]{\includegraphics[trim=0 35 0 0, height=5.5cm]{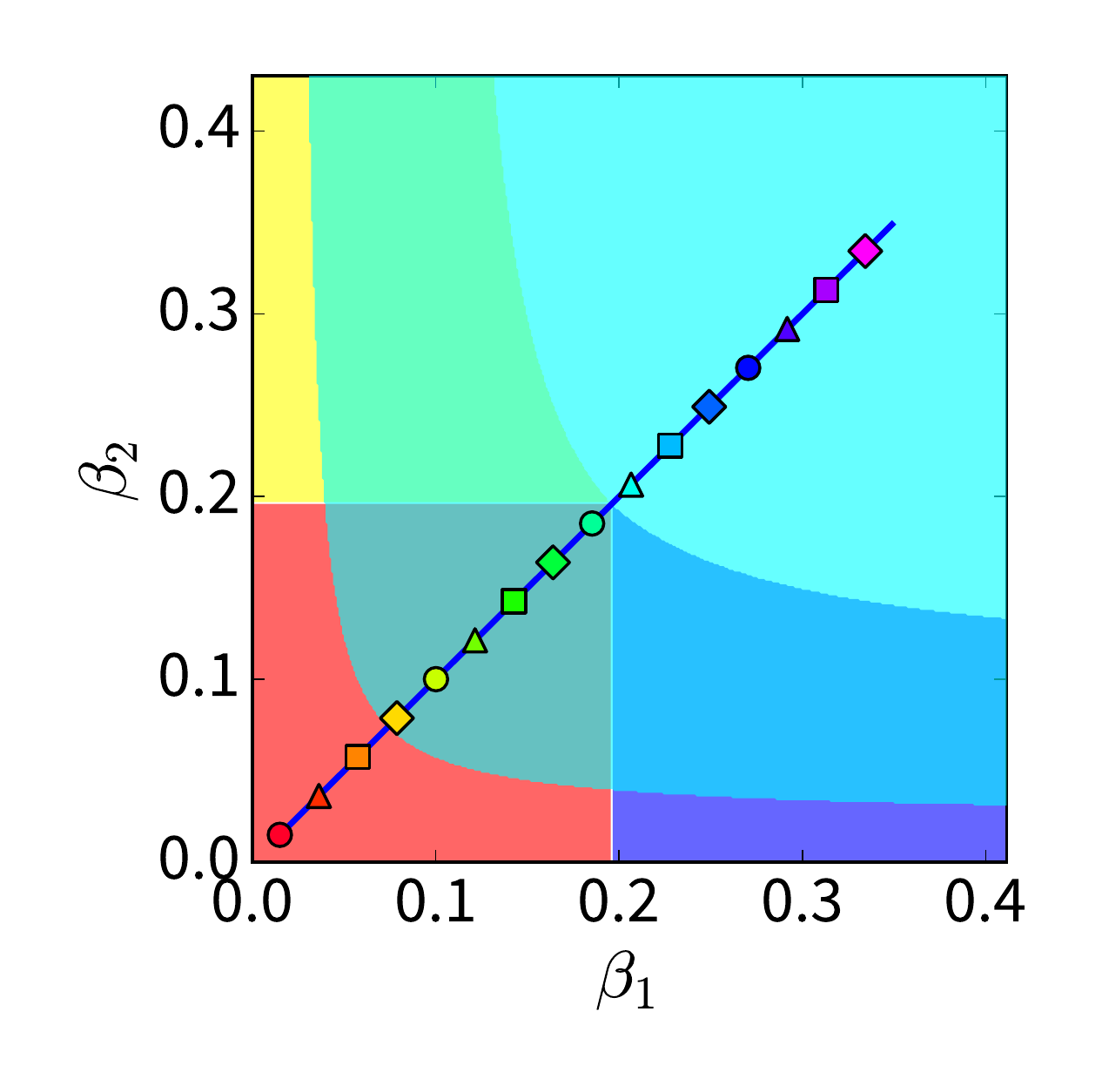}}
    \subfloat[]{\includegraphics[height=5.5cm]{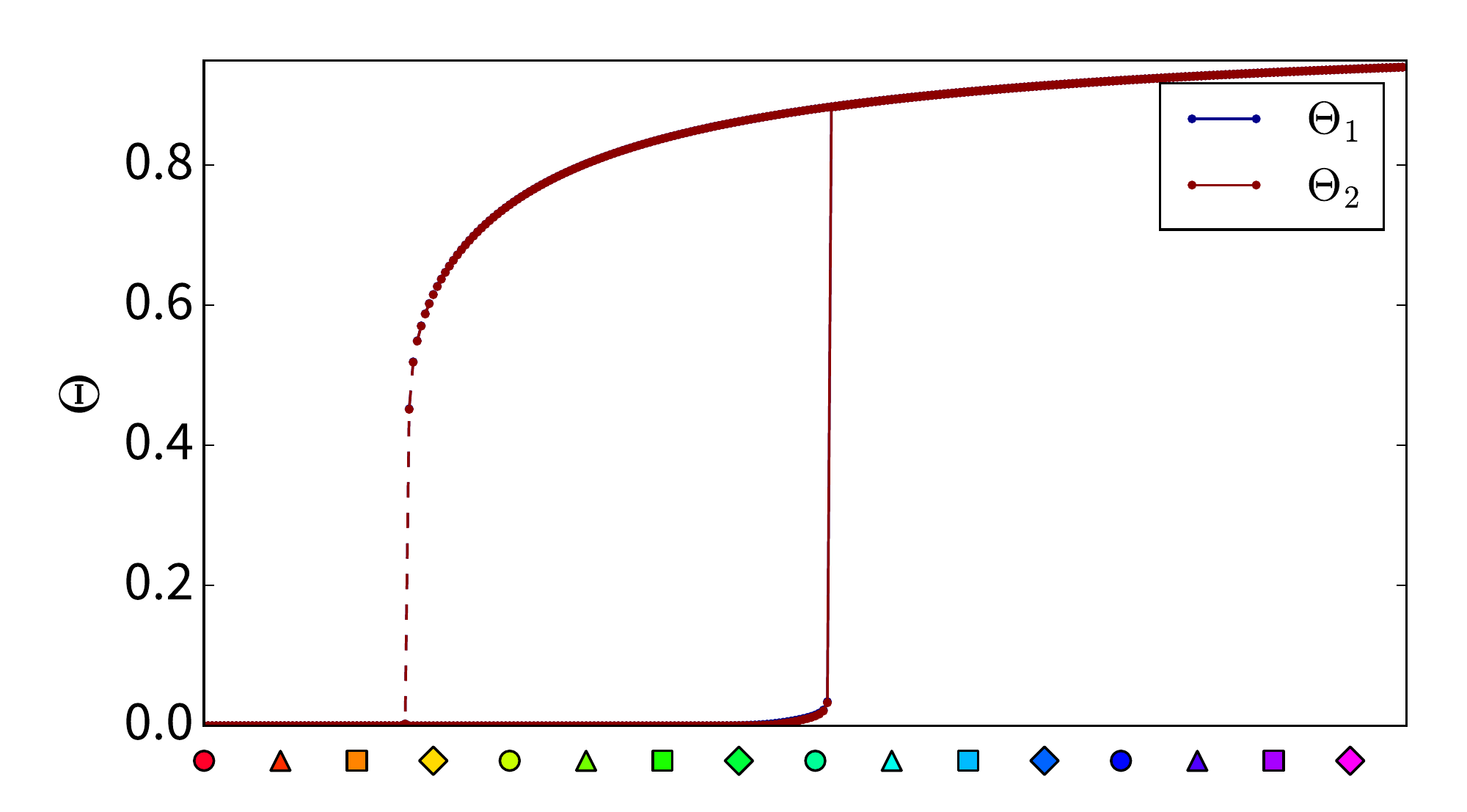}}\\
    \subfloat[]{\includegraphics[trim=0 35 0 0, height=5.5cm]{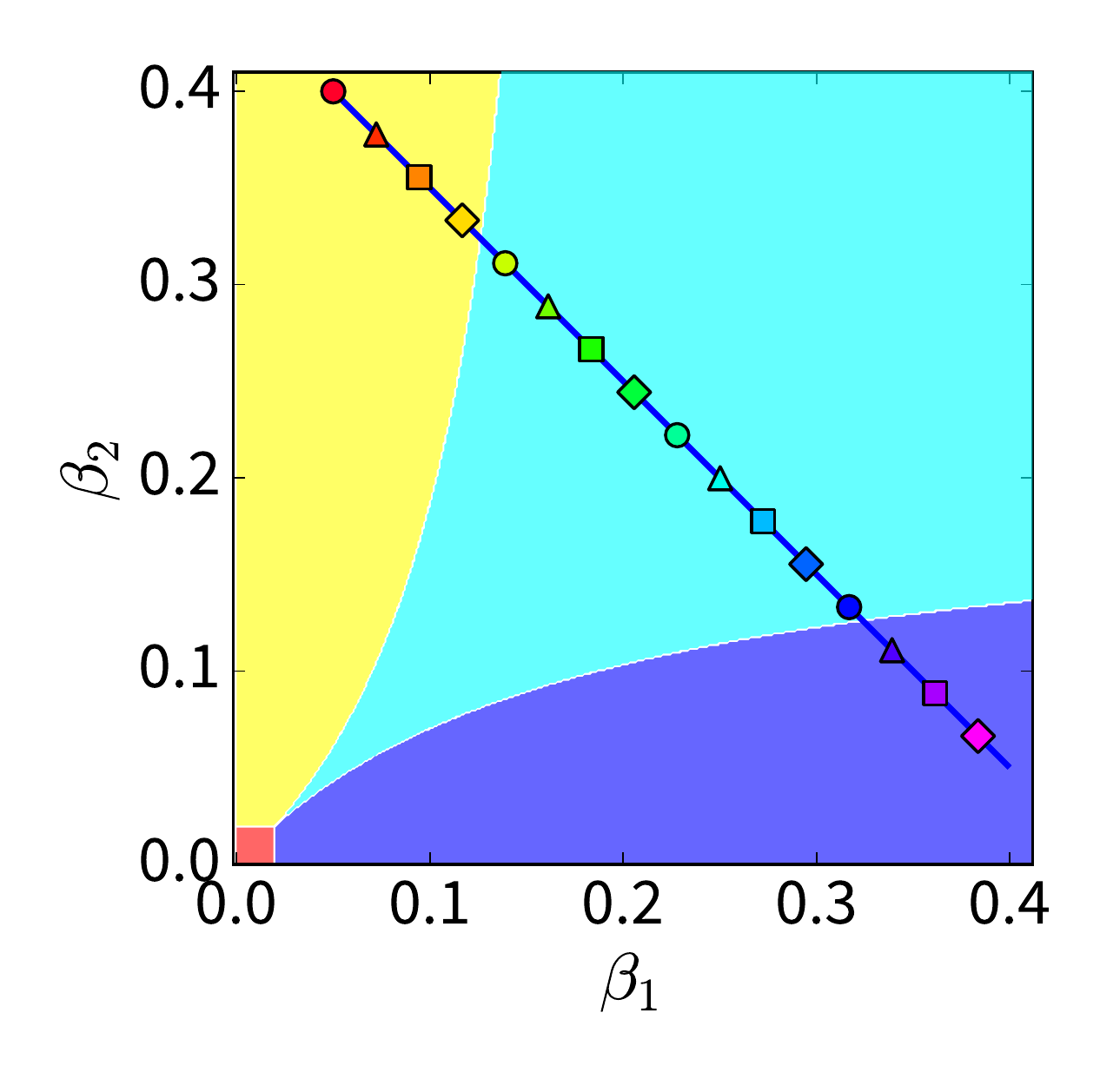}}
    \subfloat[]{\includegraphics[height=5.5cm]{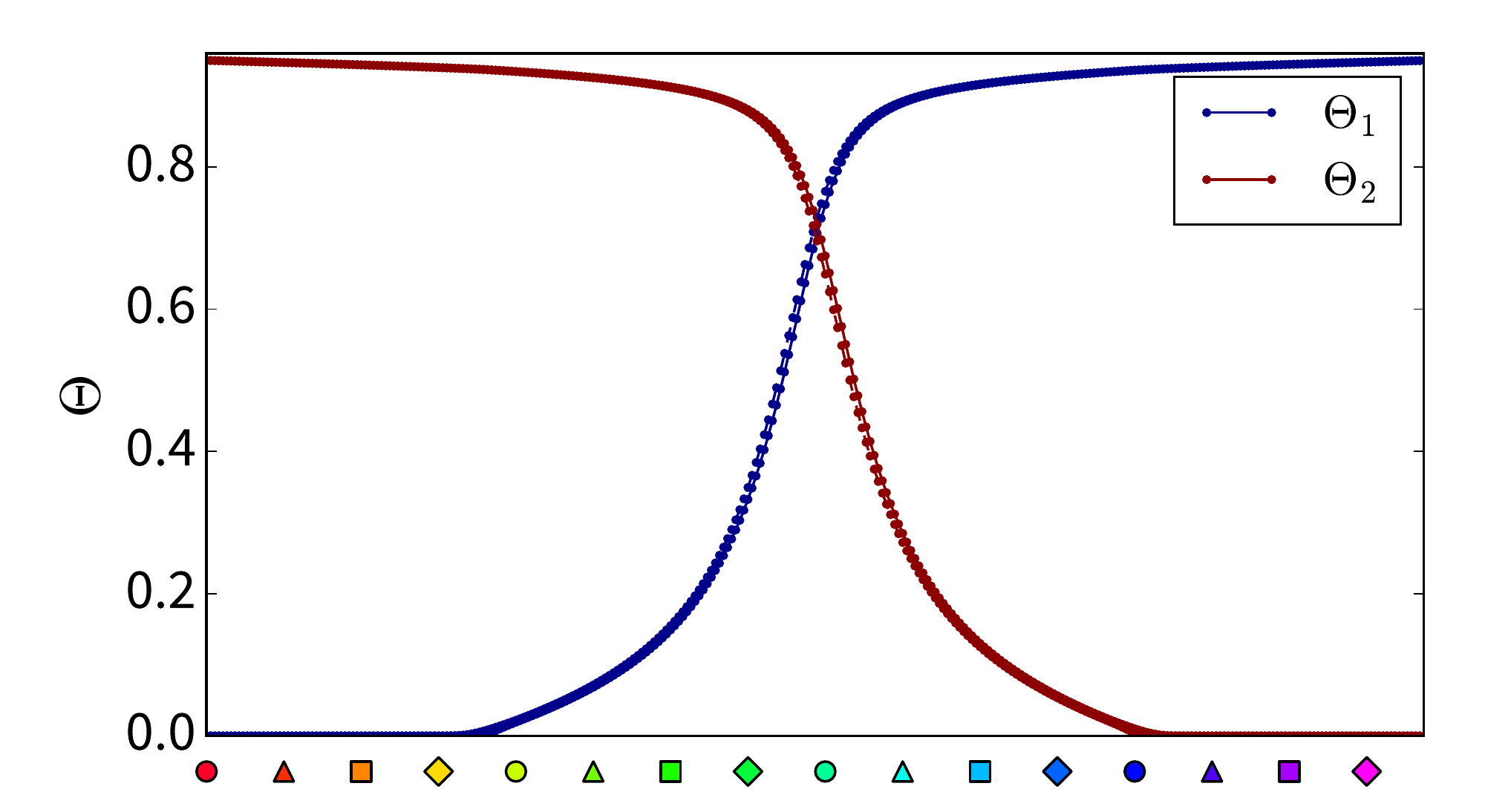}}
    \caption{{\bf Interacting epidemics} (color online). Interacting SIS dynamics with interdependent and competitive interactions.
    \textbf{(a)} Interdependent epidemic phase space. We find epidemic spread in four phases:  yellow (network 2 only), cyan (both networks), blue (network 1 only) and red (no epidemics).
    \textbf{(b)} Hysteresis. Because the cyan and red regions overlap, we find hysteresis.  This implies that an interdependent disease is harder to stop once the outbreak has occurred. 
    Blue and red curves are almost indistinguishable because the system is symmetric.
    \textbf{(c)} Competitive epidemic phase space. For an ER network with $\kk = 50$ and fraction $f = 0.9$ nodes competitive, we find epidemic spread in four phases:  yellow (network 2 only), cyan (both networks), blue (network 1 only) and red (no epidemics).
    \textbf{(d)} Coexistence. Here there is no bistability but rather non-trivial coexistence patterns of the competing diseases. Forward and backward curves are indistinguishable.
    Panels show ER network with $\kk = 50$ and fraction $f = 0.9$ nodes interdependent/competitive, simulation for $N = 1000$, each point integrated to $t=20$ and the path composed of 300 steps, in each direction.
    }
    \label{fig:sis}
\end{figure*}

\section{Epidemics}
Because the cross-system dynamic relationships defined in Eq. \eqref{eq:model} are general, it is simple to adapt them to other scenarios in which the dynamic state of a process in one network affects its activity in another layer.
In epidemics,  multiple diseases or strains can spread cooperatively~\cite{newman2013interacting,chen2013outbreaks,azimi2016cooperative}, with the onset of one disease increasing the susceptibility to other diseases. At the same time, information about vaccines can spread on social networks similar to those upon which the disease spreads, leading to competitive contagions~\cite{newman2005threshold,sanz2014dynamics,sahneh2014competitive}.  If exposure to the disease in one layer induces an individual to vaccinate in another layer, we may find that the disease layer enhances the vaccine layer, even as the vaccine layer suppresses the disease layer (the assymetric ``hybrid'' case of our model)~\cite{ahn2006epidemic}.  
 Our model of dynamic dependence offers a unified framework to study a broad spectrum of possible interactions between epidemics on multilayer networks.
For simplicity, we treat a probabilistic reversible (SIS) model which, in the single layer case evolves according to:
\begin{equation}\label{eq:modelSIS0}
\dot{x}_i=-\gamma x_i+\beta(1-x_i)\sum_{j=1}^N A_{ij}x_j,
\end{equation}
where $\gamma$ is the recovery rate and $\beta$ is the transmission rate.  For simplicity, we renormalize the time such that $\gamma = 1$ and define the local order parameter as 
\begin{equation}\label{eq:SISop}
\Theta_i = \frac{1}{\kk_i} \sum_j A_{ij} x_j.
\end{equation}
This immediately leads to the epidemics analog of Eq.\eqref{eq:model}:
\begin{equation}\label{eq:SISdep}
\dot{x}_i^\sigma=-x_i^\sigma+\beta_\sigma F_i^{\mu\to\sigma}k_i^\sigma\Theta_i^\sigma\big(1-x_i^\sigma\big),
\end{equation}
where $F_i$ is defined as in Eq. \eqref{eq:interaction} for interdependent or competitive interactions.
Using the standard mean-field theory \cite{pastorsatorras-prl2001} we let $\Theta_i \rightarrow \Theta$ and obtain the general self-consistent equations (special case of Eq. \eqref{eq:genselfconst} with $G(x,k) = \frac{x}{1 + kx}$):
\begin{equation}\label{eq:SISselfconst}
\Theta_\sigma=\frac{\beta_\sigma\Theta_\sigma F^{\mu\to\sigma}}{\langle k\rangle_\sigma}\int\limits_1^{+\infty}k^2 P_\sigma(k)\frac{ 1 }{1+\beta_\sigma k\Theta_\sigma F^{\mu\to\sigma}}\dd k.
\end{equation}
Here,  $F^{\mu\to\sigma}$ equals $\Theta_\sigma$ for the interdependent case, and $1 - \Theta_\mu$ for the competitive case.  Letting $F^{\mu\to\sigma} = 1$ we recover the known decoupled single-layer case  \cite{pastorsatorras-prl2001}.  Analagous equations for $f<1$ are provided in the Methods section.

Looking at the results of these interactions, we find a number of novel and realistic phenomena.
In the interdependent case, we find that there is a first-order transition in the forward direction and an abrupt transition in the backward direction (Fig. \ref{fig:sis}a,b) when $f<1$ and no forward transition at all when $f=0$.
In contrast to the case of synchronization, where a finite level of fluctuations (due to the quenched disorder of the natural frequency distribution) made the zero-solution spontaneously jump to the synchronized state, the system leaves the zero solution only when it becomes a proper unstable fixed point.
In practical terms, we see that for interdependent epidemics it is much harder (i.e., requires a much larger reduction in $\beta$) to stop an epidemic once it has become endemic than it is to keep it from breaking out.
Another practical consequence of the bistability is that, in contrast to non-cooperative epidemics, the zero solution has a finite basin of stability even after the endemic phase has emerged.  
This means that in cooperative/interdependent epidemics, small outbreaks are expected to die out even for comparatively high transmission rates, but that a large outbreak can become endemic.

In the competitive case we also see new behaviors that differ from synchronization.  Instead of metastability, we find a broad regime of coexistence (Fig. \ref{fig:sis}c,d).  This indicates that it is possible for the disease and its cure to coexist in the same system, presenting a challenge for any attempt to completely eradicate it.
The hybrid case (shown in Supp. Fig. 13) does not display the chaotic behavior of the analagous synchronization system but instead has oscillating but convergent attractors (Supp. Fig. 14).

\section{Connection with interdependent percolation}
We argue here that the new framework developed above represents a generalization of interdependence from percolation to general dynamic systems (those which follow Eq. \eqref{eq:0}).
The reason we describe the interaction in Eq. \eqref{eq:model} as representing a \emph{dynamic dependency link} is that it has the same impact on  \emph{dynamic order} that the percolation dependency link has on the connectivity-based order of interdependent percolation.
This is because, when we implement the dependency interaction from Eq. \eqref{eq:interaction}, the result is that as long as node $i$ is locally disordered in network ${B}$,  the coupling strength of node $i$ in network $A$ is multiplied by $|\mathpzc{z}_i^\mathrm{B}|\approx0$, effectively cutting it off from the other nodes in $A$ and suppressing the onset of order in $A$.
This reflects a dynamical generalization of interdependent percolation \cite{buldyrev-nature2010}:  in percolation the local order of a node is defined as whether any of its neighbors leads to the largest connected component; if it has no local order, then it suppresses the ability of the node that depends on it from becoming ordered.
As with interdependent percolation, dynamic interdependence leads to increased vulnerability as reflected in a higher threshold and the abruptness of the transition.
There have been several studies of competitive percolation as well, but the absence of a natural temporal evolution has led to difficulties in describing the dynamics in each system~\cite{zhao-jstatmech2013,kotnis-pre2015,watanabe-pre2016}.

One of the most intriguing findings that we report here is that all of the systems: percolation, synchronization and epidemics, undergo a remarkably similar  transition in the backward direction (from order to disorder).  In all of the cases, the transition is characterized by a slow cascading process with a ``plateau'' (Fig. \ref{fig:plateau}).
The system quickly falls from a highly ordered state to an intermediate state and then spends the bulk of its relaxation time with the order parameter maintaining an approximately constant value, before dropping to zero at an exponential rate.
While this cascade has been studied in the context of a branching process in percolation~\cite{dong-pre2014,lee-pre2016}, here we observe it through a fixed point that transitions from stable to marginally unstable (see phase flows in Supp. Fig. 2 and 4).  The macroscopic signature of this process is remarkably similar across these highly diverse processes, reflecting a universality to interdependent transitions across diverse dynamics.

\begin{figure}
    \centering
    \includegraphics[width=\columnwidth]{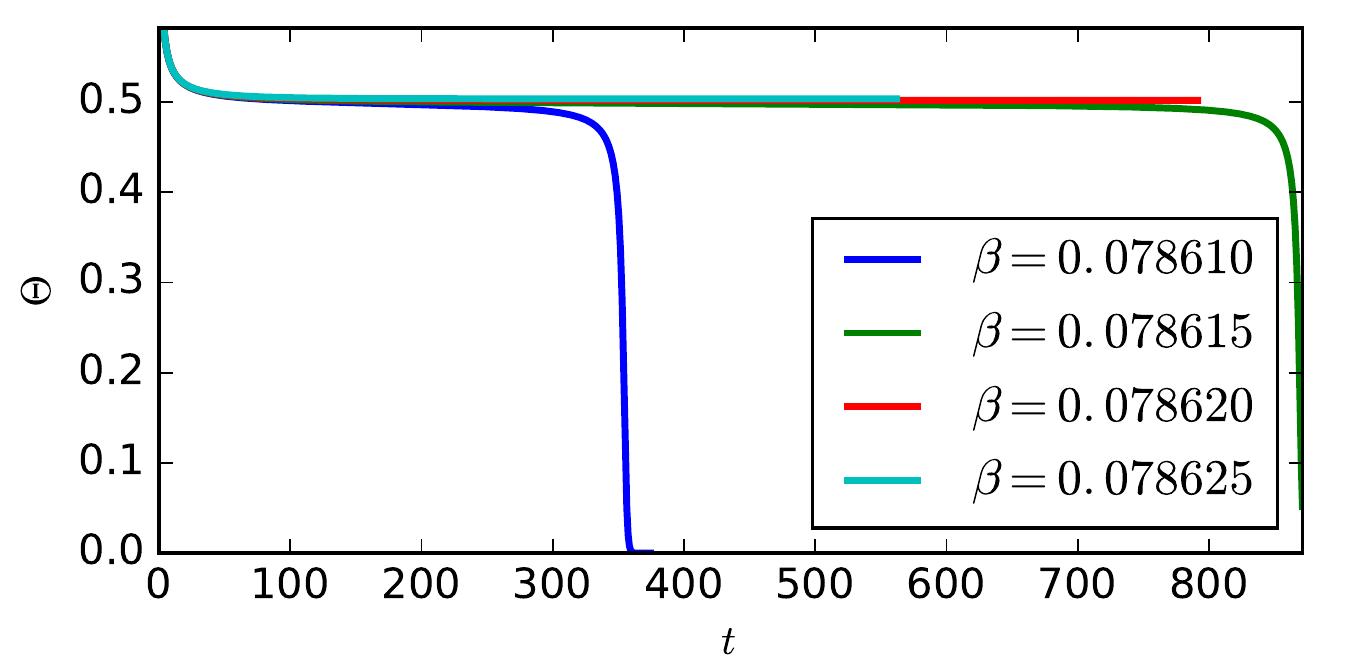}
    \caption{Hybrid transition of dynamic interdependent epidemic. As in interdependent percolation, the backwards transition (from ordered to disordered) is characterized by a ``plateau'': a steep drop in the order parameter, followed by a long almost constant period and then another steep drop. ER network with $N=1000$ and $\kk = 50$.}
    \label{fig:plateau}
\end{figure}

\section{Conclusion}
Here we have presented a universal theoretical framework for interdependent and competitive links between general dynamic systems.
Whereas previous attempts to include new effects in multi-layer systems have required ad-hoc definitions and assumptions, we now present a simple and realistic approach that can be easily adapted to a wide variety of systems.
Furthermore, because the connections are based on the fundamental statistical physical properties of local-order, standard mean-field methods can be applied straightforwardly to obtain analytic results.

This generalization of interdependence to dynamical systems allows for the development of new models for neural, social and technological systems that better capture the subtle ways in which different systems can affect one another.
Though we focus here on the fundamental interactions of interdependence and competition, more exotic interactions can be described by this framework by replacing the simple linear functions of $z_i$ in Eq.~\ref{eq:interaction} with more complex functions.
We find that the phenomenology recovers  key features of interdependent percolation but also uncovers new phenomena like oscillatory and chaotic states, metastability and coexistence, which were not known to be caused by inter-system interactions.

}

\section*{Materials \& Methods}{\small 
\paragraph*{\textbf{{\em{\small Mean-field approximation for interacting synchronization.}}}} 
We consider therefore a network of Kuramoto models, each describing a population of oscillatory elements, and model the cross-system interactions according to our framework. For simplicity, let us focus on a system of $M=2$ dynamically dependent Kuramoto models on arbitrary networks of equal size. Assuming no self-dynamical dependencies, the dynamics of the overall system is described by the set of $2N$ coupled equations 
\begin{equation}
\dot{\theta}_i^\sigma=\omega_i^\sigma+\lambda_\sigma F_i^{\mu\to\sigma}\sum_{j=1}^N A_{ij}^{(\sigma)}\sin(\theta_j^\sigma-\theta_i^\sigma),\label{eq:interKuraMM}
\end{equation}
where $i=1,\dots,N$, and $\sigma=\mathrm{A},\,\mathrm{B}$ with $\mu\neq\sigma$.

In order to identify all the possible asymptotic coherent states of the model \eqref{eq:interKuraMM}, we can follow the traditional self-consistent method developed by Kuramoto \cite{kuramoto-proceedings1975,kuramoto2012chemical}. 
In particular, let us consider the case of two uncorrelated random graphs with prescribed degree sequences \cite{bollobas2001random,newman-book2010,cohen-book2010}. 
Adopting the so-called {\em annealed network approximation} \cite{bianconi2002mean,dorogovtsev2010lectures,boccaletti-physicsreports2014}, we replace the entries $A_{ij}^{(\sigma)}$ with their ensemble averages, i.e. the probabilities $p_{ij}^{\sigma}=k_i^{\sigma}k_j^{\sigma}/(\langle k\rangle_\sigma N)$ that vertices $i$ and $j$ with degree $k_i^\sigma$ and $k_j^\sigma$, respectively, are connected in layer $\sigma=\mathrm{A},\,\mathrm{B}$, being $\langle k\rangle_\sigma$ its corresponding mean degree. 
Within this approach, each local order parameter \eqref{eq:localOP} becomes independent of the node indices, so that 
\begin{equation}\label{eq:OPKuramoto}
\mathpzc{z}_i^\sigma(t)=R_\sigma(t)e^{\imath\Psi_\sigma(t)},\qquad\forall\,i\in\mathfrak{G}_\sigma,
\end{equation}
meaning that each node feels the global mean-field measured over the entire network it belongs to. Let us further assume that in the thermodynamic limit both populations reach {\em asymptotically steady} states with constant mean-field amplitudes and constant frequencies, so that $\dot{R}_\sigma=0$ and $\ddot{\Psi}_\sigma=\dot{\Omega}_\sigma=0$ for $\sigma=\mathrm{A},\mathrm{B}$, and choose {\em equal}, symmetric and unimodal frequency distributions on layers $\mathfrak{G}_\mathrm{A}$ and $\mathfrak{G}_\mathrm{B}$. 
In this case, the equations governing the phase dynamics \eqref{eq:interKuraMM} can be rewritten as
\begin{equation}\label{eq:interKura2MM}
\dot{\varphi}_i^\sigma=\omega_i-\lambda_\sigma k_i^{\sigma}F^{\mu\to\sigma}R_{\sigma}\sin\varphi_i^\sigma,
\end{equation}
where $\varphi_i^\sigma\equiv\theta_i^\sigma-\Omega_\sigma t$ are local phase differences \cite{basnarkov2008kuramoto,petkoski2013mean,iatsenko2013stationary} defined after moving to rotating reference frames on each unit circle comoving with frequencies $\Omega_\sigma$ (see Supp. Sec. 1 for more details).\\
\indent
Solutions of the system \eqref{eq:interKura2MM}, shows that the population of oscillators in each layer splits into two groups, namely drifting and frequency locked oscillators \cite{strogatz2000kuramoto}. The phases of the latter ones are entrained by the mean-field and correspond to fixed point solutions of the system \eqref{eq:interKuraMM}, i.e. $\dot{\boldsymbol{\varphi}}^\sigma=\boldsymbol{0}$. Drifting oscillators, on the other hand, are not entrained and never reach steady states. Nonetheless, based on the model's assumptions, drifting oscillators have vanishing contributions to the complex order parameters in the thermodynamic limit (see Supp. Sec. 1), so that the overall collective behaviour of the system is dominated entirely by the locked oscillators. 
We find hence that all the possible coherent stationary states of the system \eqref{eq:interKuraMM} can be obtained by solving self-consistently a system of two integro-transcendental equations
\begin{equation}\label{eq:KuraSelfCompact}
R_\sigma=\lambda_\sigma R_\sigma\int_{\R}\mathpzc{x}\,\mathcal{I}_\sigma\big(\lambda_\sigma R_\sigma  \mathpzc{x}\big)\Gamma^{\mu\to\sigma}(\mathpzc{x})\dd\mathpzc{x},
\end{equation}
\noindent
where $\mathcal{I}_\sigma$ is an integral function accounting for the degree distributions of the network $\mathfrak{G}_\sigma$ and the natural frequency distribution of the Kuramoto oscillators sitting on its nodes (see Supp. Sec. 1), whilst $\Gamma^{\mu\to\sigma}(\mathpzc{x}):=f\delta(\mathpzc{x}-F^{\mu\to\sigma})+(1-f)\delta(\mathpzc{x}-1)$ is the distribution of the dynamical dependencies from $\mathfrak{G}_\mu$ to $\mathfrak{G}_\sigma$, with $\sigma,\,\mu=\mathrm{A},\,\mathrm{B}$ and $\mu\neq\sigma$.\\
\indent
Notice that the null solution describing the mutual incoherent phase always satisfies Eqs.~\eqref{eq:KuraSelfCompact}, though it not always stable.
Non-vanishing solutions of \eqref{eq:KuraSelfCompact} can be found numerically for general graph topologies and frequency distributions \cite{lafuerza2010nonuniversal,omel2012nonuniversal}, once the strategy for the dynamical interactions between the layers has been chosen.

{
\paragraph*{\textbf{{\em{\small Mean-field approximation for interacting epidemics.}}}} 
Consistently with the general framework \eqref{eq:localOP}--\eqref{eq:model}, let us introduce a global order parameter for SIS epidemics, here defined as $\Theta:=\sum_{i} k_i\Theta_i/\sum_i k_i$, where
$$
\Theta_i:=\frac{1}{k_i}\sum_{j=1}^N A_{ij}x_j,
$$
are local order parameters measuring the local spread of the disease in the neighbourhoods of the $i^{\mathrm{th}}$ node \cite{barrat-book2008,newman-book2010,pastorsatorras-revmodphys2015}. Inserting the latter into the above equations, one finds 
$$
\dot{x}_i=-\gamma x_i+\beta(1-x_i)k_i\Theta_i.
$$
Stationary solutions can then be found by setting $\dot{x}_i=0$ in the latter expression, which yields the local self-consistent equations
$$
x_i=\frac{\beta k_i \Theta_i}{\gamma+\beta k_i\Theta_i},
$$
for every $i=1,\dots,N$. The above identifies what is in the literature known as individual-based mean field approach. In what follows, we will consider instead the degree-based approximation introduced by by Pastor-Satorras et {al.} \cite{pastorsatorras-prl2001,pastorsatorras-revmodphys2015}, which allows us to solve just one self-consistent equation for the global order parameter. Moreover, it is straightforward to prove that this approximation is actually equivalent to the annealed network approximation for uncorrelated random graphs \cite{newman-book2010,pastorsatorras-revmodphys2015}, in which case the local order parameters $\Theta_i$ become node-independent so that, $\Theta\equiv\Theta_i=\frac{1}{\langle k\rangle N}\sum_{i=1}^N k_jx_j$. Stationarity implies then 
$$
x_i=\frac{\beta k_i\Theta}{\gamma +\beta k_i\Theta},
$$
so that one can find all the possibly stationary states by solving the self-consistent equation 
\begin{displaymath}
\Theta=\beta\Theta\int\limits_1^{+\infty}
\frac{k^2 P(k)}{\langle k\rangle\left(\gamma+\beta k\Theta\right)}\dd k,
\end{displaymath}
where we replaced the sums with definite integrals over the degree distribution after moving to the thermodynamic limit. 
Moving to dynamically dependent epidemic processes, let us consider the the moment the case of fully interdependency links between the replica nodes. In such case, node $i$ in layer $\mathfrak{G}_\mathrm{A}$ feels the local spread of the disease of its replica node in layer $\mathfrak{G}_\mathrm{B}$, which adaptively changes its local rate of transmission, i.e. $\beta_\mathrm{A}\mapsto\beta_\mathrm{A}\Theta_i^{\mathrm{A}}$ and viceversa. Hence 
\begin{displaymath}
\begin{cases}
\begin{aligned}
\dot{x}_i^{\mathrm{A}}&\,=-\gamma_\mathrm{A}x_i^{\mathrm{A}}+\beta_\mathrm{A}\Theta_i^{\mathrm{B}}\big(1-x_i^{\mathrm{A}}\big)k_i^{\mathrm{A}}\Theta_i^{\mathrm{A}},\\
\dot{x}_i^{\mathrm{B}}&\,=-\gamma_\mathrm{B}x_i^{\mathrm{B}}+\beta_\mathrm{B}\Theta_i^{\mathrm{A}}\big(1-x_i^{\mathrm{B}}\big)k_i^{\mathrm{B}}\Theta_i^{\mathrm{B}},
\end{aligned}
\end{cases}
\end{displaymath}
and one arrives eventually to the following self-consistent system of equations
\begin{displaymath}
\Theta_\sigma=\frac{\beta_\sigma}{\langle k\rangle_\sigma}\int\limits_1^{+\infty}k^2P_\sigma(k)\frac{\Theta_\sigma \Theta_\mu}{\gamma_\sigma+\beta_\sigma k\Theta_\sigma\Theta_\mu}\dd k,
\end{displaymath}
for $\sigma=\mathrm{A},\,\mathrm{B}$ and $\mu\neq\sigma$.\\}

}

\bibliography{NoN.bib}

\end{document}